\documentclass[conference]{IEEEtran}

\IEEEoverridecommandlockouts
\usepackage{amsmath,amssymb,amsfonts}
\usepackage{algorithmic}
\usepackage[ruled]{algorithm2e}
\usepackage{array}
\usepackage{textcomp}
\usepackage{stfloats}
\usepackage{url}
\usepackage{color}
\usepackage{verbatim}
\usepackage{graphicx}
\usepackage{tabularx}
\usepackage{multirow}
\usepackage{array}
\usepackage{booktabs}
\usepackage{graphicx}
\usepackage{parskip}

\newtheorem{definition}{Definition}
\usepackage[numbers]{natbib}
\usepackage{bm}
\usepackage{mathtools}

\usepackage{subfigure}

\usepackage{bibspacing}
\graphicspath{ {./fig/} }
\usepackage{enumitem}
\def\BibTeX{{\rm B\kern-.05em{\sc i\kern-.025em b}\kern-.08em
    T\kern-.1667em\lower.7ex\hbox{E}\kern-.125emX}}
\begin{document}

\title{Social and Physical Attributes-Defined Trust Evaluation for Effective Collaborator Selection \\ in Human-Device Coexistence Systems\\
}

\author{\IEEEauthorblockN{Botao Zhu and Xianbin Wang}
\IEEEauthorblockA{Dept. of Electrical and Computer Engineering, Western University,
London, Ontario N6A 3K7 CANADA \\}   

}

\maketitle

\begin{abstract}
In human-device coexistence systems, collaborations among devices are determined by not only physical attributes such as network topology but also social attributes among human users. Consequently, trust evaluation of potential collaborators based on these multifaceted attributes becomes critical for ensuring the eventual outcome. However, due to the high heterogeneity and complexity of physical and social attributes, efficiently integrating them for accurate trust evaluation remains challenging. To overcome this difficulty, a canonical correlation analysis-enhanced hypergraph self-supervised learning (HSLCCA) method is proposed in this research. First, by treating all attributes as relationships among connected devices, a relationship hypergraph is constructed to comprehensively capture inter-device relationships across three dimensions: spatial attribute-related, device attribute-related, and social attribute-related. Next, a self-supervised learning framework is developed to integrate these multi-dimensional relationships and generate device embeddings enriched with relational semantics. In this learning framework, the relationship hypergraph is augmented into two distinct views to enhance semantic information. A parameter-sharing hypergraph neural network is then utilized to learn device embeddings from both views. To further enhance embedding quality, a CCA approach is applied, allowing the comparison of data between the two views. Finally, the trustworthiness of devices is calculated based on the learned device embeddings. Extensive experiments demonstrate that the proposed HSLCCA method significantly outperforms the baseline algorithm in effectively identifying trusted devices. 

\end{abstract}


\vspace{-0.08 in}
\section{Introduction}
Human-device coexistence systems feature a deep integration of human social attributes with intelligent devices, enabling them to act as socially connected entities, not just computational units~\cite{M. A. Díaz}. Human social network structures govern device deployment, connection strategies, and collaborator selection, further shaping system operation and performance. As a result, the connectivity and collaboration between devices are not only influenced by physical networks and resources, but also by social relationships among human users. For example, in the metaverse, two friends attending the same virtual concert from different physical locations can have their VR headsets establish a temporary, prioritized connection to optimize their shared experience.

In such systems, the selection of collaborators significantly affects task execution efficiency, cooperation stability, and overall system performance. Trust, as a core metric for evaluating the reliability and risk level of collaborators, is becoming essential for ensuring effective collaboration. From the perspective of task completion, trust is defined as the ability of a collaborator to fulfill the task owner's specific task requirements~\cite{chain_of_trsut}. In existing research, some studies assess device trustworthiness by leveraging social attributes, such as interest similarity, through predefined rules or models~\cite{P. Dong}. Others focus on historical device behaviors, such as task success rates and communication stability, employing machine learning for trust evaluation~\cite{D. Zhang}. However, these traditional trust evaluation models fail to deliver accurate results in human-device coexistence systems, as they are incapable of effectively capturing the complex relationships in both physical and social domains. To achieve accurate trust evaluation in such systems, the design of trust evaluation models should consider the following two key aspects:
i) Multi-source trust fusion: trust evaluation models should integrate not only a device's physical capabilities and physical behavioral performance in collaboration but also its social attributes. For instance, a device possessing significant computational power might still be excluded from task allocation if it demonstrates limited social connectivity with other entities due to potential risks; ii) Semantics-enhanced representation: trust evaluation models must be capable of semantically understanding and representing both social and physical attributes to facilitate more accurate trust assessment. 





Therefore, this study proposes a canonical correlation analysis-enhanced hypergraph self-supervised learning (HSLCCA) method to semantically represent and integrate both social and physical attributes for accurate trust evaluation in human-device coexistence systems. The main contributions of this paper are summarized as follows.
\begin{itemize}[leftmargin=*]
    \item We are the first to propose integrating devices' social and physical attributes for trust evaluation. By treating all attributes as relationships connecting devices, we propose a novel hypergraph-based approach to comprehensively capture multi-dimensional inter-device relationships.

    \item  A self-supervised learning framework is proposed to fuse these multi-dimensional relationships and generate device embeddings enriched with complex semantic information. In this learning framework, the relationship hypergraph is augmented into two distinct views to enhance semantic diversity. A parameter-sharing hypergraph neural network (HGNN) is then applied to both views to learn informative device embeddings.

    \item  To further improve the quality of embeddings, a CCA approach is employed to compare the data from both views, optimizing the model's learning performance. Finally, the trustworthiness of devices is computed based on the learned device embeddings. 
    
\end{itemize}

\section{System Model and Problem Definition}
\label{problem}
A human-device coexistence system comprising $\bm{A} = \{a_1,\dots, a_I\}$ devices is considered. Devices are connected through various attributes, and each attribute is regarded as a relationship connecting devices. Let $\bm{S}$ denote the set of all relationships. To ensure reliable collaboration, each device must evaluate the trustworthiness of potential collaborators based on these diverse relationships before selecting the most trustworthy one. Accordingly, trust is formally defined as follows:
\vspace{-0.04 in}
\begin{definition}[Relationship-defined trust]
   \textit{For any pair of devices $a_i, a_j \in \bm{A}$, the trust of device $a_i$ in $a_j$ is a measure of $a_i$’s confidence in the reliability of $a_j$ based on a set of relationships $\bm{S}$, which is given by}
   \vspace{-0.01 in}
   \begin{align}
       T_{a_i \to a_j} = TRUST(a_i,a_j,\bm{A}, \bm{S}). 
   \end{align}
\end{definition}
\vspace{-0.09 in}

According to Definition 1, trust evaluation between two devices requires not only considering their direct relationships but also accounting for the potential influence of other devices and relationships within the system. Therefore, accurate trust evaluation requires a system-level perspective that captures the influence of all interconnected relationships. The goal of trust evaluation is to identify the most reliable collaborator for a task initiator. If $a_i$ is the task initiator, the problem of selecting the most trusted collaborator can be expressed as follows:
\vspace{-0.05 in}
\begin{align}
    \arg \max_{a_j \in \bm{A}, a_j \neq a_i} T_{a_i \to a_j}.
\end{align}
To address this problem, the HSLCCA algorithm is proposed to facilitate comprehensive relationship representation, effective fusion of multi-dimensional relationships, and accurate trust evaluation. Based on the evaluation results, the most reliable collaborator can be efficiently identified.

\section{Trust Evaluation via CCA-Enhanced Hypergraph Self-Supervised Learning}
Due to the complexity and diversity of relationships, an effective tool is required to represent these relationships, thereby enabling accurate trust evaluation. Graphs are the most commonly used tool to represent relationships. However, they can only capture low-order point-to-point relationships between devices, such as the topology connectivity relationship, and are unable to capture high-order relationships between multiple devices, such as the common interest relationship. Hypergraphs are gaining attention as a powerful tool for modeling relationships among multiple nodes, including both low-order and high-order relationships~\cite{11096939}. A hypergraph $\mathcal{H} = (\mathcal{A}, \mathcal{E})$ consists of a set of nodes $\mathcal{A}$ and a set of hyperedges $\mathcal{E}$. Each hyperedge $e \in \mathcal{E}$ connects a group of nodes, capturing the relationship among them. A node $a \in \mathcal{A}$ can establish various types of relationships with other nodes within the system. The hypergraph structure can be represented by an incidence matrix $H \in \{0,1 \}^{|\mathcal{A}| \times |\mathcal{E}|}$, where each entry $h_{ae} = 1$ if $a \in e$ and $h_{ae} = 0$ otherwise~\cite{11000000000}.


Leveraging the advantages of hypergraphs, this section presents HSLCCA, a hypergraph-based trust evaluation approach, for human-device coexistence systems. The proposed approach comprises three primary components: i) Hypergraph-driven relationship representation: Hypergraphs are used to capture relationships among devices, resulting in a relationship hypergraph $\mathcal{H}^{\text{all}}$; 
ii) CCA-enhanced hypergraph self-supervised learning: This process involves fusing multi-dimensional relationships and generating device embeddings enriched with relational semantics; 
iii) Trust computation: Trust between devices is calculated based on their embeddings.

\subsection{Hypergraph-Driven Relationship Representation}

To comprehensively uncover relationships, we divide them into three categories: spatial attribute-related, device attribute-related, and social attribute-related. In the following sections, we will discuss how to represent each category using hypergraphs.

\vspace{-0.05 in}
\textit{1) Spatial attribute-Related}: Spatial attribute-related relationships describe the spatial distribution and connectivity of devices within a physical space. These relationships encompass attributes related to network topology and physical space. For example, in large-scale systems, devices located within the same physical area are more likely to have direct network links and shorter communication paths, resulting in lower latency and more stable interactions. Devices that ensure reliable interactions typically exhibit higher trustworthiness. Therefore, incorporating these relationships into trust evaluation is essential. In this study, the spatial attribute-related relationships include the network connectivity relationship $s^{\text{net}} \in \bm{S}$ and the physical proximity relationship $s^{\text{phy}} \in \bm{S}$.


\vspace{-0.03 in}
\textbf{Network connectivity relationship}: $s^{\text{net}}$ indicates whether a direct communication link exists between any two devices, as directly connected devices have a higher potential for collaboration. If there is a communication link between devices $a_i$ and $a_j$, we represent this connection with a hyperedge $e^{\text{net}}_{a_i,a_j} = \{a_i,a_j\}$. By transforming all direct communication links into hyperedges, a network connectivity relationship hypergraph can be obtained, denoted as $\mathcal{H}^{\text{net}} = \{\dots, e^{\text{net}}_{a_i,a_j}, \dots \}$. 

\textbf{Physical proximity relationship}: $s^{\text{phy}}$ reflects the relative positions of devices in physical space. Devices that are geographically closer often experience lower transmission delays and more stable communication, which positively influences collaborative efficiency. To effectively capture this relationship, a clustering technique is employed to partition devices based on their physical locations. Through iterative refinement, it can group nearby devices into the same cluster while distinguishing those that are spatially distant. This clustering process can be represented as:
\vspace{-0.05 in}
\begin{align}
    \{c_1,\dots,c_K \} = Clustering(\bm{A}),
\end{align}
where $c_k, k = 1, \dots, K$ represents a cluster. Subsequently, $c_k$ is encapsulated by an edge to form a hyperedge $e_k^{\text{phy}}$. Ultimately, all clusters are transformed into hyperedges, collectively forming a physical proximity relationship hypergraph $\mathcal{H}^{\text{phy}}$. 

\vspace{-0.08 in}
\textit{2) Device Attribute-Related}: Device attribute-based relationships reflect the relevance among devices based on their exhibited characteristics, such as resource configurations, functional roles, and behavioral patterns. For example, devices from the same manufacturer often share similar resource profiles and operational behaviors. Likewise, devices with similar functions are more likely to collaborate for task execution. In this category, two primary relationships are considered: the collaboration relationship $s^{\text{his}} \in \bm{S}$ and the resource similarity relationship $s^{\text{res}} \in \bm{S}$.


\vspace{-0.06 in}
\textbf{Collaboration relationship}: $s^{\text{his}}$ captures the historical cooperation performance among devices, reflecting their efficiency and compatibility. Stable and effective historical collaborations indicate strong interoperability between devices, thereby increasing the likelihood of future cooperation. If a group of devices $a_i$, $a_j$, and $a_m$ previously collaborated to complete a task $t$, a hyperedge $e^{\text{his}}_t$ is created to encapsulate these three devices, representing the collaboration. The weight of the hyperedge indicates the effectiveness of the collaboration, where 1 denotes a successful collaboration and 0 denotes a failed collaboration. All historical collaborations are converted into corresponding hyperedges to form the collaboration relationship hypergraph $\mathcal{H}^{\text{his}} = \{\dots, e^{\text{his}}_t, \dots \}$. 

\vspace{-0.06 in}
\textbf{Resource similarity relationship}: $s^{\text{res}}$ indicates the similarity of devices in terms of computing power, storage, and other aspects. Devices with similar resources tend to exhibit similar operating modes and behavioral characteristics. Since there are many types of devices in the system, for simplification, we categorize them based on basic device types, such as smartphone, computer, etc. Assuming that $a_i$, $a_j$, and $a_m$ are smartphones, a hyperedge $e^{\text{res}}$ can be used to encapsulate these three devices, indicating that they belong to the same device type. Ultimately, all such hyperedges representing different device types can be obtained, collectively forming a resource similarity relationship hypergraph $\mathcal{H}^{\text{res}}$.

\vspace{-0.01 in}
\textit{3) Social Attribute-Related}:
The above two categories explore the relationships between devices from a physical perspective. However, collaboration between devices extends beyond these relationships, as they also influence each other through complex social attributes. 
For example, devices may build trust through common partners, forming social groups or circles. Therefore, social attributes are essential for comprehensive trust evaluation. In this study, two social attribute-based relationships are considered: the common interest relationship $s^{\text{int}} \in \bm{S}$ and the common friend relationship $s^{\text{fri}} \in \bm{S}$. 

\vspace{-0.05 in}
\textbf{Common interest relationship}: $s^{\text{int}}$ refers to the similarity between devices in terms of specific functions, use cases, or user preferences, such as health monitoring or security surveillance. Devices with common interests are more likely to collaborate effectively. Assuming the system contains a total of $B$ interests, each interest $b \in B$ can be represented as a hyperedge, denoted by $e^{\text{int}}_{b}$, that encapsulates all devices sharing that interest. Ultimately, the collection of all hyperedges forms a common interest relationship hypergraph $\mathcal{H}^{\text{int}} = \{\dots, e^{\text{int}}_{b}, \dots\}$.

 \begin{figure}[t!]
	\centering
	  \includegraphics[scale=0.9]{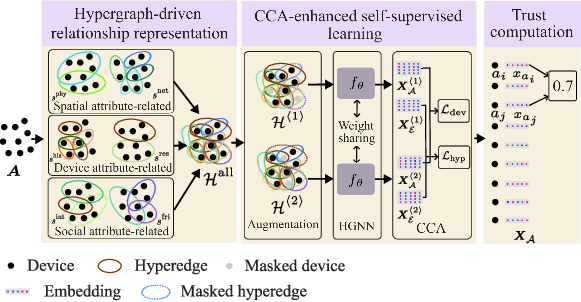}
	\caption{The proposed HSLCCA method, including relationship representation, fusion, and trust computation.}
	\label{hslcca}
\end{figure}

\vspace{-0.07 in}
\textbf{Common friend relationship}: $s^{\text{fri}}$ refers to the connection between two devices established by sharing the same friend.  For instance, if device $a_i$ maintains connections with both devices $a_j$ and $a_m$, then $a_j$ and $a_m$ are considered to have a common friend relationship through $a_i$. This relationship can be represented by a hyperedge $e^{\text{fri}}_{a_j,a_m}$ that connects the indirectly linked devices. The set of all such hyperedges constitutes the common friend relationship hypergraph, denoted as $\mathcal{H}^{\text{fri}} = \{\dots,  e^{\text{fri}}_{a_j,a_m}, \dots\}$.


To integrate all relationships, the obtained hypergraphs are combined to generate a new relationship hypergraph $\mathcal{H}^{\text{all}} = \mathcal{H}^{\text{net}} \cup \mathcal{H}^{\text{phy}} \cup \mathcal{H}^{\text{his}} \cup \mathcal{H}^{\text{res}} \cup \mathcal{H}^{\text{int}} \cup \mathcal{H}^{\text{fri}} = \left(\mathcal{A}, \mathcal{E} \right)$, where $\mathcal{A} = \bm{A}$, and $\mathcal{E}$ is the set of all hyperedges. For consistency in notation, all hyperedges are re-expressed as $\mathcal{E} = \{e_n \}_{n=1}^{|\mathcal{E}|}$. Each hyperedge is associated with a weight $w_n$, and the matrix of weights is $\bm{W} \in \mathbb{R}^{|\mathcal{E}| \times |\mathcal{E}|}$. The feature matrix of all devices in $\mathcal{H}$ is represented as $\bm{X}_{\mathcal{A}} \in \mathbb{R}^{|\mathcal{A}| \times d}$, and the incidence matrix of $\mathcal{H}$ is given by $\bm{H} \in \mathbb{R}^{|\mathcal{A}| \times |\mathcal{E}|}$. The degree of devices is denoted by the diagonal matrix $\bm{D}_{a} \in \mathbb{R}^{|\mathcal{A}| \times |\mathcal{A}|}$, where each element $\delta(a_i) = \sum_{e_n \in \mathcal{E}} w_{n} h(a_i,e_n)$. The degree of hyperedges is denoted by the diagonal matrix $\bm{D}_{e} \in \mathbb{R}^{|\mathcal{E}| \times |\mathcal{E}|}$, where each element $\delta(e_n) = \sum_{a_i \in e_n}h(a_i,e_n)$ representing the number of devices connected by $e_n$.

\vspace{-0.08 in}
\subsection{CCA-Enhanced Hypergraph Self-Supervised Learning}

Although $\mathcal{H}^{\text{all}}$ captures all relationships between devices, trust between any pair of devices cannot be directly calculated from these relationships. To address this issue, devices and their relationships need to be mapped into the same dimensional space. Hypergraph learning can achieve this goal, which uses a function  $f_\theta: \mathcal{H}^{\text{all}} \to (\bm{X}_{\mathcal{A}}, \bm{X}_{\mathcal{E}})$ to learn the devices' embeddings $\bm{X}_{\mathcal{A}}$ and hyperedges' embeddings $\bm{X}_{\mathcal{E}}$ within the same dimensional space. To train $f_\theta$, we use the CCA-enhanced self-supervised learning approach. The core idea is to generate two augmented views from the raw data, offering different contexts or semantics, and then train a model through self-supervised learning to maximize the agreement between the two views. Therefore, the model training process comprises three steps: 1) Two data views are generated through hypergraph augmentation; 2) Hypergraph embedding is performed using HGNN; 3) The model is trained via CCA-enhanced self-supervised learning.

\textit{1) Hypergraph Augmentation}: 
To create two augmented views, two types of data augmentation are utilized: device masking and membership masking. In device masking, devices are randomly masked with a certain probability, which is given by
\vspace{-0.2 in}
\begin{align}
    Aug(\mathcal{A}, p^{\mathcal{A}}) = \mathcal{A} \odot \bm{M}^{\mathcal{A}},
\end{align}
where $\odot$ represents element-wise multiplication, $\bm{M}^{\mathcal{A}}$ is the mask matrix in which each element is independently sampled from a Bernoulli distribution $\mathcal{B}(1 - p^{\mathcal{A}})$, and $p^{\mathcal{A}}$ is the probability of devices being masked. Membership masking refers to randomly removing some node-hyperedge memberships from the hypergraph, which is formulated as
\begin{align}
    Aug(\bm{H}, p^{\bm{H}}) = \bm{H} \odot \bm{M}^{\bm{H}},
\end{align}
where $\bm{M}^{\bm{H}}$ is the masking matrix whose elements are sampled from a Bernoulli distribution $\mathcal{B}(1 - p^{\bm{H}})$, and $p^{\bm{H}}$ is the masking probability. Finally, we can obtain two views of $\mathcal{H}^{\text{all}}$
\begin{align}
    \mathcal{H}^{\langle 1 \rangle} &= (Aug(\mathcal{A}, p^{\mathcal{A}}), Aug(\bm{H}, p^{\bm{H}})), \\
    \mathcal{H}^{\langle 2 \rangle} &= (Aug(\mathcal{A}, p^{\mathcal{A}}), Aug(\bm{H}, p^{\bm{H}})).
\end{align}
\textit{2) Hypergraph Embedding via HGNN}: To generate device and hyperedge embeddings for the two augmented views, a parameter-sharing HGNN is employed. Each view is fed into the HGNN, which applies a two-stage neighborhood aggregation process: device-to-hyperedge and hyperedge-to-device. HGNN iteratively updates the representation of each hyperedge by aggregating representations of its incident devices, which is given by
\begin{align}
\label{etoa}
    \bm{x}^{(l)}_{e_n} = f^{(l)}_{\mathcal{A} \to \mathcal{E}} \left(\bm{x}^{(l-1)}_{e_n}, \left \{ \bm{x}^{{(l-1)}}_{a_i} : a_i \in e_n \right \}\right),
\end{align}
where $\bm{x}^{(l-1)}_{e_n}$ and $\bm{x}^{{(l-1)}}_{a_i}$ are the embeddings of $e_n$ and $a_i$ at layer $(l - 1)$, respectively. In addition, the representation of each device is updated iteratively through aggregating representations of its incident hyperedges
\begin{align}
    \label{atoe}
    \bm{x}^{(l)}_{a_i} = f^{(l)}_{\mathcal{E} \to \mathcal{A}} \left(\bm{x}^{(l - 1)}_{a_i}, \left \{ \bm{x}^{(l)}_{e_n}: a_i \in e_n \right \} \right).
\end{align}
Formally, equations (\ref{etoa}) and (\ref{atoe}) in the $l$-th layer of HGNN can be represented in matrix form
\begin{align}
    \bm{X}^{(l)}_{\mathcal{E}} &=  \phi\left (\bm{D}^{-1}_{e} \bm{H}^T 
    \bm{X}^{(l - 1)}_{\mathcal{A}} \bm{\Theta}^{(l)}_{\mathcal{E}} \right ), \\
    \bm{X}^{(l)}_{\mathcal{A}} &= \phi \left (\bm{D}^{-1}_a \bm{H} \bm{W} \bm{X}^{(l)}_{\mathcal{E}} \bm{\Theta}^{(l)}_{\mathcal{A}} \right), 
\end{align}
where $\bm{X}^{(l)}_{\mathcal{E}} \in \mathbb{R}^{|\mathcal{E}| \times d}$ and $\bm{X}^{(l)}_{\mathcal{A}} \in \mathbb{R}^{|\mathcal{A}| \times d}$ are hyperedge and device embeddings at the $l$-th layer. $d$ is the embedding dimensionality. $ \bm{\Theta}^{(l)}_{\mathcal{E}}$ and $\bm{\Theta}^{(l)}_{\mathcal{A}}$ are trainable parameters for $f^{(l)}_{\mathcal{A} \to \mathcal{E}}$ and $f^{(l)}_{\mathcal{E} \to \mathcal{A}}$, respectively. $\phi(\cdot)$ denotes a nonlinear activation function. Finally, HGNN outputs the device and hyperedge embeddings for each augmented view, denoted as $(\bm{X}^{{\langle 1 \rangle}}_{\mathcal{A}}, \bm{X}^{{\langle 1 \rangle}}_{\mathcal{E}})$ for $\mathcal{H}^{\langle 1 \rangle}$ and $(\bm{X}^{{\langle 2 \rangle}}_{\mathcal{A}}, \bm{X}^{{\langle 2 \rangle}}_{\mathcal{E}})$ for $\mathcal{H}^{\langle 2 \rangle}$.

\textit{3) CCA-Enhanced Self-Supervised Learning}: To obtain more meaningful device embeddings,  it is necessary to perform comparison and self-supervised learning on the two augmented views. Traditional contrastive self-supervised learning methods require a large number of negative samples, resulting in high computational overhead and efficiency challenges. CCA has gained increasing attention in multi-view learning due to its capability to capture both linear and non-linear relationships between views without the need for negative sampling, thereby significantly reducing computational complexity. As a multivariate analysis method, CCA aims to maximize the correlation between data representations from different views, enhancing the effectiveness of multi-view learning models~\cite{H. Zhang}. Therefore, we use CCA to design two optimization objectives for self-supervised learning to maximize the agreement between two augmented views. 

\textbf{Device-level CCA optimization objective}: The device embeddings are first normalized along the instance dimension such that each feature dimension has zero mean and a standard deviation of $1/\sqrt{|\mathcal{A}|}$
\vspace{-0.06 in}
\begin{align}
    \widehat{\bm{X}}_{\mathcal{A}}^{\langle v \rangle} = \frac{\bm{X}_{\mathcal{A}}^{\langle v \rangle} - \mu (\bm{X}_{\mathcal{A}}^{\langle v \rangle}) }{\sigma(\bm{X}_{\mathcal{A}}^{\langle v \rangle}) \sqrt{|\mathcal{A}|}}, v = 1, 2, 
\end{align}
 where $\mu(\cdot)$ and $\sigma(\cdot)$ denote computing mean value and standard deviation, respectively. Then, we construct the device-level CCA loss function 
 \begin{align}
   \label{device_cca}
     \mathcal{L}_{\text{dev}} = \mathcal{L}^{\text{inv}}_{\text{dev}} + \lambda_{\text{dev}} \mathcal{L}^{\text{dec}}_{\text{dev}}, 
 \end{align}
where $\lambda_{\text{dev}}$ is a non-negative hyperparameter trading off two terms, and $\mathcal{L}^{\text{inv}}_{\text{dev}}$ and $\mathcal{L}^{\text{dec}}_{\text{dev}}$ are the invariance term and the decorrelation term, respectively. $\mathcal{L}^{\text{inv}}_{\text{dev}}$ preserves the device-wise invariant information, which is given by
\begin{align}
    \mathcal{L}^{\text{inv}}_{\text{dev}} = \parallel \widehat{\bm{X}}_{\mathcal{A}}^{\langle 1 \rangle} - \widehat{\bm{X}}_{\mathcal{A}}^{\langle 2 \rangle}  \parallel^2_{F},
\end{align}
where $\parallel \cdot \parallel^2_{F}$ represents the square of Frobenius norm. $\mathcal{L}^{\text{dec}}_{\text{dev}}$ is used to prevent dimension collapse, which is defined as
\begin{small}
    \begin{align}
   \hspace{-0.1 in} \mathcal{L}^{\text{dec}}_{\text{dev}} = \parallel {(\widehat{\bm{X}}_{\mathcal{A}}^{\langle 1 \rangle})}^\top \widehat{\bm{X}}_{\mathcal{A}}^{\langle 1 \rangle} - \bm{I} \parallel^2_{F} + \parallel {(\widehat{\bm{X}}_{\mathcal{A}}^{\langle 2 \rangle})}^\top \widehat{\bm{X}}_{\mathcal{A}}^{\langle 2 \rangle} - \bm{I} \parallel^2_{F},
\end{align}
\end{small}
\hspace{-0.04 in}where $\bm{I} \in \mathbb{R}^{d \times d}$ is the identity matrix. 

\textbf{Hyperedge-level CCA optimization objective}: This aims to distinguish the representations of the same hyperedge across the two augmented views from other hyperedges, helping the model retain hyperedge-wise information. Similar to the device-level optimization objective, the hyperedge-level optimization objective can be formulated by the following equations
\vspace{-0.15 in}
\begin{equation}
   \widehat{\bm{X}}_{\mathcal{E}}^{\langle v \rangle} = \frac{\bm{X}_{\mathcal{E}}^{\langle v \rangle} - \mu (\bm{X}_{\mathcal{E}}^{\langle v \rangle}) }{\sigma(\bm{X}_{\mathcal{E}}^{\langle v \rangle}) \sqrt{|\mathcal{E}|}}, v = 1, 2,
\end{equation}
\begin{equation}
    \label{hyperedge_cca}
    \mathcal{L}_{\text{hyp}} = \mathcal{L}^{\text{inv}}_{\text{hyp}} + \lambda_{\text{hyp}} \mathcal{L}^{\text{dec}}_{\text{hyp}},
\end{equation}
\begin{equation}
        \mathcal{L}^{\text{inv}}_{\text{hyp}} = \parallel \widehat{\bm{X}}_{\mathcal{E}}^{\langle 1 \rangle} - \widehat{\bm{X}}_{\mathcal{E}}^{\langle 2 \rangle}  \parallel^2_{F}, 
\end{equation}
\vspace{-0.2 in}
\begin{small}
\begin{align}
   \mathcal{L}^{\text{dec}}_{\text{hyp}} &= \parallel {(\widehat{\bm{X}}_{\mathcal{E}}^{\langle 1 \rangle})}^\top \widehat{\bm{X}}_{\mathcal{E}}^{\langle 1 \rangle} - \bm{I} \parallel^2_{F} + \parallel {(\widehat{\bm{X}}_{\mathcal{E}}^{\langle 2 \rangle})}^\top \widehat{\bm{X}}_{\mathcal{E}}^{\langle 2 \rangle} - \bm{I} \parallel^2_{F},
\end{align}
\end{small}
\hspace{-0.05 in}where $\mathcal{L}_{\text{hyp}}$, $\mathcal{L}^{\text{inv}}_{\text{hyp}}$, and $\mathcal{L}^{\text{dec}}_{\text{hyp}}$ are the hyperedge-level CCA loss, hyperedge-wise invariant term, and hyperedge-wise decorrelation term, respectively. By integrating (\ref{device_cca}) and (\ref{hyperedge_cca}), the total loss is formulated as
\vspace{-0.06 in}
\begin{align}
    \mathcal{L} = \mathcal{L}_{\text{dev}} + \lambda_1 \mathcal{L}_{\text{hyp}} + \lambda_2 \parallel \Theta \parallel^2,
\end{align}
where $\lambda_1$ and $\lambda_2$ are the weights of $\mathcal{L}_{\text{hyp}}$ and $L_2$ regularization, respectively. $\Theta$ is the set of learnable parameters in HGNN. The proposed HSLCCA jointly optimizes device-level CCA loss and hyperedge-level CCA loss, which enables the learned embeddings of devices and hyperedges to preserve both the device- and hyperedge-level structural information.

\vspace{-0.025 in}
\subsection{Trust Computation}

After obtaining the embeddings of devices, their trust values can be directly calculated based on these embeddings, as the inter-device relationships have been fused within the embedding space. The trust of device $a_i$ in device $a_j$ is obtained by calculating the similarity between $a_i$'s embedding $\bm{x}_{a_i}$ and $a_j$'s embedding $\bm{x}_{a_j}$
\vspace{-0.04 in}
\begin{align}
  T_{a_i \to a_j}  = \frac{\bm{x}_{a_i} \cdot \bm{x}_{a_j}}{\parallel \bm{x}_{a_i}\parallel \parallel \bm{x}_{a_j} \parallel}, \bm{x}_{a_i}, \bm{x}_{a_j} \in \bm{X}_{\mathcal{A}}.
\end{align}
Finally, $a_i$ evaluates the trust values of all potential collaborators and selects the one with the highest trust value as the trusted collaborator.

\vspace{-0.07 in}
\section{Experiments}
\label{simulation}


\subsection{Experimental Setup}
\textit{1) Dataset}: To assess the effectiveness of the proposed HSLCCA method, we utilize the Sigcomm-2009 dataset \cite{b20}. The dataset includes rich information associated with devices or users, such as friendships, shared interests, activities, and message logs. It comprises 76 nodes and 18,226 interaction records collected over a period of four days. As the original dataset lacks geographic coordinates for the nodes, we assign each node a position in a two-dimensional space. Because the dataset is generated by only one type of phone, we define three distinct phone types and randomly assign one type to each node.

\textit{2) Hyperparameters}: Both $p^{\bm{H}}$ and $p^{\bm{\mathcal{A}}}$ are set to 0.5. $\lambda_{\text{dev}}$ is set to 0.0002, and  $\lambda_{\text{hyp}}$ is set to 0.0035~\cite{X. Yang}. The weights $\lambda_1$ and $\lambda_2$ are assigned values of 1 and 0.05, respectively. The embedding size is set to 512. HGNN is trained using the Adam optimizer with a weight decay of $10^{-5}$. The proposed model is implemented by Python and PyTorch.

\vspace{-0.02 in}
\subsection{Comparison of t-SNE Visualization of Device Embeddings}

\begin{figure}[!]
\centering
\includegraphics[scale=0.9]{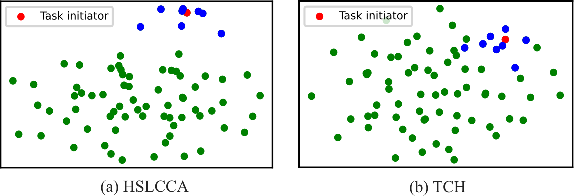}
\caption{Comparison of t-SNE visualization of the device embeddings produced by HSLCCA and TCH. (The red node represents the task initiator, the blue points are the top 8 nodes with the highest trust values evaluated by the task initiator, and the green points represent other nodes.)}
\label{tnse}
\end{figure}

Since the device embeddings generated by the proposed HSLCCA contain complex relationships, we visually compare these embeddings with those generated by TCH~\cite{b2} in this subsection. All high-dimensional embeddings are projected into a two-dimensional space using the t-SNE tool, which preserves the relative proximity of similar data points during dimensionality reduction, enabling a more intuitive visualization of the differences. Node 5 is chosen as the task initiator, and the top 8 nodes with the highest trust values evaluated by the task initiator are marked in blue. As shown in Fig.~\ref{tnse}, compared with the TCH method, the HSLCCA method brings these top trusted nodes closer to the task initiator, forming a compact cluster. This indicates that the HSLCCA method more effectively captures the structure and semantic proximity in the relationships. 

To quantitatively assess clustering quality, the silhouette score (SS) is employed. An SS value close to 1 indicates well-separated and compact clusters, reflecting strong intra-cluster cohesion and clear inter-cluster distinction. Conversely, a negative SS value suggests that nodes are incorrectly assigned. The SS value obtained by HSLCCA is 0.41, while the TCH method yields a value of 0.01. This significant difference demonstrates that HSLCCA excels at distinguishing the most trusted nodes from the others, leading to more accurate trust evaluation.

\begin{figure}[!t]
\centering
\includegraphics[scale=0.75]{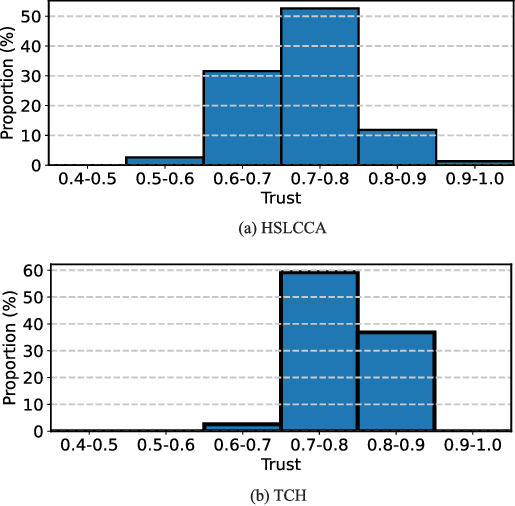}
\caption{Comparison of the trust value distribution of nodes.}
\label{bar}
\end{figure}


Fig.~\ref{bar} presents a comparison of the trust value distribution of nodes, where the horizontal axis denotes trust value intervals and the vertical axis indicates the proportion of nodes within each interval. It is observed that the TCH method yields high trust values predominantly in the (0.8, 0.9] interval. In contrast, the HSLCCA method identifies a small number of nodes with trust values in the higher (0.9, 1.0] interval. This result further highlights the effectiveness of HSLCCA in distinguishing the most trusted nodes.

\vspace{-0.1 in}
\subsection{Sensitivity Analysis of $p^{\bm{H}}$ and $p^{\mathcal{A}}$}

\begin{figure}[!]
\centering
\includegraphics[scale=0.87]{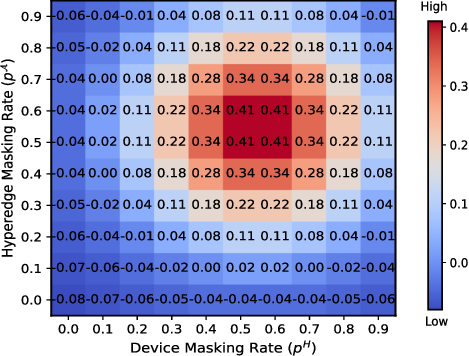}
\caption{Impact of $p^\mathcal{A}$ and $p^{\bm{H}}$ on SS.}
\label{heatmap}
\end{figure}

Fig.~\ref{heatmap} shows the impact of $p^{\bm{H}}$ and $p^{\mathcal{A}}$, with each grid value representing a SS value. As we can see, when both $p^{\mathcal{A}}$ and $p^{\bm{H}}$ fall within the range of 0.4 to 0.7, the SS value is relatively high, indicating that nodes with high trust values can be effectively identified. When $p^{\mathcal{A}}$ and $p^{\bm{H}}$ are small, the two views become similar, which is insufficient to capture the discriminant ability of HGNN. Conversely, when $p^{\mathcal{A}}$ and $p^{\bm{H}}$ are large, the underlying semantics of the original relationship hypergraph are broken, resulting in ineffective identification of trusted nodes. 

\vspace{-0.1 in}
\subsection{Comparison of selected trustworthy nodes when changing the number of nodes}
\begin{figure}[!]
\centering
\includegraphics[scale=0.56]{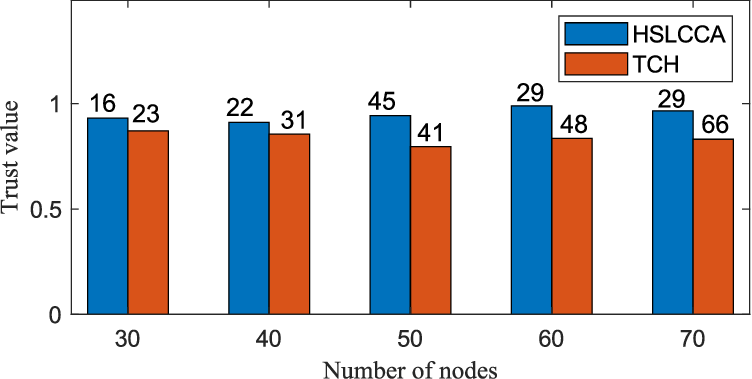}
\caption{Comparison of selected trustworthy nodes.}
\label{changsize}
\end{figure}

In this subsection, the influence of the total number of nodes on the selection of the most trusted node is examined. As shown in Fig.~\ref{changsize}, the x-axis represents the total number of nodes in the system, while the numbers above the bars indicate the identifiers of the selected most trusted nodes. For example, when the total number of nodes is 30, the HSLCCA method identifies node 16 as the most trusted, with a trust value of 0.93. As the number of nodes increases, the most trusted nodes selected by both methods also vary. Nevertheless, the nodes selected by HSLCCA consistently exhibit the highest trust values. Furthermore, node 29 is selected as the most trusted node at both 60 and 70 nodes by HSLCCA, despite slight fluctuations in its trust value. This indicates that an increase in the number of nodes enriches the social relationship structure, thereby influencing the trust evaluation outcomes.

\vspace{-0.08 in}
\section{Conclusion}
\label{conclusion}

To ensure reliable collaboration in human-device coexistence systems, this paper proposes the HSLCCA approach, which integrates attribute representation, fusion, and trust evaluation to identify trustworthy collaborators. First, hypergraphs are employed to explore and represent the complex relationships among devices from three dimensions. Subsequently, the CCA-enhanced hypergraph self-supervised learning is used to fuse and learn these complex relationships, ultimately incorporating latent relationship information into the device embeddings. Finally, trust values between devices are determined based on their embeddings. Extensive experiments demonstrated that the proposed HSLCCA method can effectively distinguish between trusted and untrusted nodes, and consistently select the most trusted nodes compared to the baseline algorithm.



\end{document}